\documentclass[twocolumn,aps,showpacs,prb,tightenlines,amsmath,amssymb]{revtex4}
\usepackage{bm}
\usepackage{graphicx}
\usepackage{amssymb}               
\usepackage{amsmath}
\usepackage{colordvi}
\usepackage{calrsfs} 
\newcommand{\bgreek}[1]{\mbox{\boldmath$#1$\unboldmath}}
\begin{document} 

\title{Electron spin diffusion and transport in graphene}
 
\author{P. Zhang}
\author{M. W. Wu}
\thanks{Author to whom correspondence should be addressed}
\email{mwwu@ustc.edu.cn.}
\affiliation{Hefei National Laboratory for Physical Sciences at
  Microscale and Department of Physics, 
University of Science and Technology of China, Hefei,
  Anhui, 230026, China}
\date{\today}

\begin{abstract} 

We investigate the spin diffusion and transport in a graphene
monolayer on SiO$_2$ substrate by means of the microscopic kinetic
spin Bloch equation approach. The substrate causes a strong Rashba spin-orbit coupling
field $\sim 0.15$~meV, which might be accounted for by the impurities
initially present in the substrate or even the substrate-induced structure distortion. By surface chemical doping with Au atoms, this Rashba
spin-orbit coupling is further strengthened as the adatoms can
distort the graphene lattice from $sp^2$ to $sp^3$ bonding structure. By fitting the Au doping
dependence of spin relaxation from Pi {\sl et al.}
[Phys. Rev. Lett. {\bf 104}, 187201 (2010)], the Rashba spin-orbit coupling
coefficient is found to increase approximately linearly from 0.15 to
0.23~meV with the increase of Au density. With this strong spin-orbit
coupling, the spin diffusion or transport length is 
comparable with the experimental values. In the strong scattering
limit (dominated by the electron-impurity scattering in our study), the spin
diffusion is uniquely determined by the Rashba
spin-orbit coupling strength and insensitive to the temperature,
electron density as well as scattering. With the presence of an electric field along the spin
injection direction, the spin transport length can be modulated by
either the electric field or the electron density. It is shown
that the spin diffusion and transport show an anisotropy with respect to the
polarization direction of injected spins. The spin diffusion or transport lengths
with the injected spins polarized in the plane defined by the
spin-injection direction and the direction perpendicular to the
graphene are identical, but
longer than that with the injected spins polarized vertical to
this plane. This anisotropy differs from 
the one given by the two-component drift-diffusion model, which 
indicates equal spin diffusion or transport lengths when the injected spins are polarized in
the graphene plane and relatively shorter lengths when the injected spins are polarized perpendicular
to the graphene plane.

\end{abstract}
\pacs{72.25.Rb, 75.40.Gb, 72.80.Vp, 71.70.Ej}

\maketitle

\section{Introduction}
Graphene is considered to be a promising candidate for the spintronic
applications recently,\cite{Tombros_07,Geim_07,castro,hwang,novoselov,
Novoselov_Science,bolotin,Loss_hyper,Kane_SOC,
Hernando_06,Min,Fabian_SOC,jozsa_08,goto,hill,cho,haugen,wang,ohishi,brey,gusynin,kane} partly due to the perfect two dimensionality,
gate-voltage-tunable charge carrier type and density,\cite{hwang,castro} high
mobility\cite{novoselov,Novoselov_Science,bolotin,morozov,du} as well as the 
potentially long spin relaxation time limited by the small intrinsic
spin-orbit and hyperfine couplings.\cite{yzhou,Loss_hyper,Kane_SOC,
Hernando_06,Min,Fabian_SOC,abde} From the high mobility and long spin relaxation
time, a long spin relaxation length, favorable to the spin information transport and
manipulation, is anticipated. However, both the spin relaxation time and transport length were
experimentally found to be much smaller than
expected.\cite{Tombros_07,han,Popinciuc,Tombros_08,Jozsa_09,Pi,Fabian_SR,yzhou} This
suggests that the spin relaxation in the  
experiments is most likely to be contributed by extrinsic factors such
as the possible impurity-enhanced spin relaxation\cite{Popinciuc,Jozsa_09}
 via the Elliot-Yafet\cite{ey}
  mechanism or the enhanced Rashba spin-orbit coupling field\cite{Kane_SOC,Min}
from the impurities.\cite{Castro_imp,varykhalov,abde} The
  former case may exist in a highly dirty graphene sample and causes the
spin relaxation time $\tau_s$ to be proportional to the momentum 
relaxation time $\tau_p$.\cite{Popinciuc,Jozsa_09} However, for the
latter case, the Dyakonov-Perel (DP) spin relaxation 
mechanism\cite{dp} dominates and
the relation $\tau_s\propto\tau_p$ is absent. In fact, recently
  Pi {\sl et al.} reported that $\tau_s$ increases
with decreasing $\tau_p$ in the surface chemical doping experiment
with Au atoms on graphene,\cite{Pi} indicating that the DP spin 
relaxation mechanism is important there.
However, the relation $\tau_s\propto 1/\tau_p$, valid when
the DP spin relaxation mechanism is
dominant and the scattering is strong enough, 
is not obeyed in their experiment.\cite{Pi} Nevertheless, we
 will show that this deviation can be understood by taking account of
the strengthening of the Rashba spin-orbit coupling with the
increasing coverage of Au adatoms. The Rashba spin-orbit
  coupling, referred to as an extrinsic one, is due to the broken of
  the inversion symmetry which can be caused by either a perpendicular electric
  field, the interaction with substrate, or the atoms adsorbed on the
  surface.\cite{Kane_SOC,Castro_imp,Min,varykhalov,abde} The
  contribution of the electric field to the Rashba spin-orbit coupling is small ($\sim
  {\mu}$eV under a perpendicular electric field as large as 1~V/nm),
  \cite{Fabian_SR,abde} while the adatoms can effectively enhance the Rashba spin-orbit
  coupling to be of order of 10~meV by distorting the graphene lattice
  from $sp^2$ to $sp^3$ bonding structure.\cite{abde,varykhalov,Castro_imp}

In this work, we investigate the spin diffusion and transport limited
by the DP mechanism in a graphene monolayer on SiO$_2$ substrate as presented by Pi {\sl et
  al.}.\cite{Pi} To account for the short spin relaxation time
($\sim$70~ps) before Au doping in the experiment,\cite{Pi} we assume that the impurities inevitably present in
  the substrate, as well as the other effects such as the
  substrate-induced structure distortion, cause a strong Rashba
  spin-orbit coupling. When the surface chemical doping by Au atoms\cite{Pi} is performed,
the Rashba spin-orbit coupling coefficient $\alpha_{\rm R}$ is
further strengthened. By fitting the chemical doping dependence of spin
  relaxation time from the experimental data,\cite{Pi} we obtain the chemical
  doping dependence of $\alpha_{\rm R}$. It is found that
  $\alpha_{\rm R}$ increases approximately linearly with the density of adatoms
  when the latter is not too high. With this essential information
  obtained, we then study the spin diffusion and transport in the
  graphene layer. The method utilized in our study is the kinetic spin Bloch
equation (KSBE) approach which has been successfully applied to the 
study of spin dynamics in semiconductors.\cite{wu-review} In the framework of this 
approach, the spatial spin precession frequency during the
steady-state scattering-free spin diffusion (assumed to be along the
$x$-axis)
is\cite{wengtran,chengtrans1,chengtrans2,zhangtrans,wu-review}
\begin{equation}
 {\bgreek \omega}_{\bf  k}=(2\bm{\Omega}_{\bf k}+g\mu_B{\bf B})/\partial_{k_x}\varepsilon_{\bf
    k}.
\end{equation}
Here $\bm{\Omega}_{\bf k}$ is the DP term, ${\bf B}$ is the 
external magnetic field and $\varepsilon_{\bf k}$ is the electron energy
spectrum. The momentum dependence of ${\bgreek \omega}_{\bf k}$ leads to 
the inhomogeneous broadening in spin precession, with which any
scattering (including the Coulomb scattering) can cause an irreversible
spin relaxation along with spin diffusion and transport.\cite{wengtran,chengtrans1,chengtrans2,zhangtrans,wu-review}
It is noted that different DP terms as well as different energy spectra lead to
distinct momentum dependences of ${\bgreek \omega}_{\bf k}$. For graphene, $\varepsilon_{\bf
  k}=\hbar v_{\rm F}k$ with $v_{\rm F}=10^6$~m/s being the Fermi velocity and
\begin{equation}
  \bm{\Omega}_{\bf  k}=\alpha_{\rm R}(-\sin\theta_{\bf k},
 \cos\theta_{\bf k},0)
\label{omega}
\end{equation} 
with $\theta_{\bf k}$ being the polar angle of
  momentum ${\bf k}$. Therefore in the absence of any external
  magnetic field
\begin{equation} 
  {\bgreek \omega}_{\bf k}=2\alpha_{\rm R}(-\tan\theta_{\bf k}, 1,0)/(\hbar
    v_{\rm F}),
  \label{frequency}
  \end{equation}
which depends on the angle $\theta_{\bf k}$ rather than the magnitude of
$k$.  This indicates that the inhomogeneous broadening 
  is {\em insensitive} to temperature and electron density as long as
  $\alpha_{\rm R}$ is fixed. Therefore the spin diffusion is only
  possible to be modulated effectively by the
  scattering.\cite{prepare} However, in this work it is revealed
  that when the scattering is strong enough (just as in the graphene layer 
  under study), the spin diffusion becomes insensitive to the
  scattering. As a result, the spin diffusion is uniquely
  determined by $\alpha_{\rm R}$. Moreover, the mean spin
  precession frequency $\langle {\bgreek
    \omega}_{\bf  k}\rangle=\frac{2\alpha_{\rm R}}{\hbar v_{\rm F}}(0,
  1, 0)$ shows a strong anisotropy which can also lead 
to the anisotropy of spin diffusion
  with respect to the spin polarization direction. This anisotropy is
  found to be quite different from the widely believed one predicted
  from the two-component drift-diffusion model.\cite{dd,yu,fabiandd,fabiandd1,fabiandd2} The discrepancy reveals the
  inadequacy of the two-component drift-diffusion model, especially for the cases
  with spin precession in spatial domain.

This paper is organized as follows. In Sec.~II, we present the model
and introduce the KSBEs. In Sec.~III, we first investigate the spin relaxation by fitting the
experimental data from Pi {\sl et al.}\cite{Pi} to obtain essential
parameters and then study the spin diffusion and transport in graphene. Both the analytical and numerical investigations are
performed. By comparing the results from the analytical and numerical
studies, we find that the analytical model depicts the zero-electric-field spin diffusion perfectly and the
nonzero-electric-field spin transport with a small discrepancy which
increases with the strength of the electric field. At last we summarize in Sec.~IV.

\section{Model and KSBEs}
The $n$-doped graphene monolayer under investigation is on a SiO$_2$
substrate perpendicular to the $z$-axis. The depth of the SiO$_2$ substrate is assumed to be
$a=300$~nm and the electric field along the $z$-axis is $E_z=V_g/a$ with $V_g$ being the gate
voltage. The spins are injected at $x=0$ and diffuse or transport along the $x$-axis. The external electric field, if 
applied, is along the
$x$-axis, i.e., ${\bf E}=E{\bf\hat x}$. Under the basis laid
out  in Refs.~\onlinecite{yzhou} and 
\onlinecite{Fabian_SR}, the Hamiltonian of electrons can be 
written as\cite{yzhou} 
\begin{eqnarray}
  \nonumber
  H &=& \sum_{{\mu}{\bf k}ss^\prime}
  \big[(\varepsilon_{\bf k}-\lambda_{\rm I}
  +eEx)  \delta_{ss^\prime}
  +\bm{\Omega}_{\bf k}\cdot{\bgreek\sigma}_{ss^\prime}\big] {c_{\mu{\bf k}s}}^{\dagger}
  c_{\mu{\bf k}s^\prime} \\ && \mbox{}+ {H}_{\rm int}.
  \label{H_eff}
\end{eqnarray}
Here $\mu$ labels the valley located at $K$ or $K^\prime$
point, ${\bgreek\sigma}$ denote the Pauli matrices and $c_{\mu{\bf k}s}$ (${c_{\mu{\bf k}s}}^{\dagger}$) is
the annihilation (creation) operator of electron in $\mu$ valley
with momentum ${\bf k}$ (relative to the valley center) and spin $s$ ($s=\pm \frac{1}{2}$). $\lambda_{\rm I}$ is the
intrinsic spin-orbit coupling constant and $-e$ is the electron charge ($e>0$).
The coefficient in the Rashba term $\bm{\Omega}_{\bf
  k}$ [Eq.~(\ref{omega})] reads
 $\alpha_{\rm R}=\zeta E_z+\eta$, with the first term
contributed by the electric field along the $z$-axis and the
second term by the substrate (including the impurities initially present inside) as well as the adatoms from surface chemical doping. The coefficient $\zeta$ is $5\times
10^{-3}$~meV$\cdot$nm/V (Refs.~\onlinecite{abde} and \onlinecite{Fabian_SR}). The 
Hamiltonian $H_{\rm int}$ consists of the electron-impurity, 
electron-phonon as well as electron-electron Coulomb
interactions.\cite{yzhou} We adopt the minimal model proposed by Adam
and Das Sarma\cite{Adam_08} to depict the electron-impurity
scattering. Within this model, only the intravalley electron-impurity scattering is
  important while the intervalley electron-impurity scattering is negligible
  due to the large momentum transfer from one valley to the other and
  the finite distance between the impurity layer and the graphene
  plane. The intravalley electron-impurity scattering matrix element reads
$|U_{{\bf k}-{\bf k^\prime}}|^2=N_i|V_{{\bf k}-{\bf
    k^\prime}}|^2e^{-2d|{\bf k}-{\bf k^\prime}|}$,\cite{Adam_08} where $N_i$ is the
effective impurity density, $d$ is the effective distance of impurities
from the graphene layer\cite{Adam_08} and $V_{{\bf k}-{\bf k^\prime}}$ is the
Coulomb potential under the random phase
approximation.\cite{ramezanali} The electron-phonon
  scattering includes the intravalley electron-acoustic phonon
  scattering,\cite{hwang1} the intervalley and intravalley
  electron-optical phonon scattering,\cite{lazzeri}
  as well as the intravalley electron-optical surface phonon
  scattering.\cite{fratini} The electron-electron Coulomb scattering
  includes both the intervalley and intravalley scattering, with the
  screening under random phase approximation given in Ref.~\onlinecite{ramezanali}.

By using the nonequilibrium Green function method,\cite{Haug_1998}
the KSBEs are constructed as\cite{wu-review}
\begin{eqnarray}\nonumber
  \frac{\partial \rho_{\mu\bf k}(x,t)}{\partial t}&=&\left.\frac{\partial\rho_{\mu\bf k}(x,t)}{\partial t}\right|_{\rm
    dri}+\left.\frac{\partial\rho_{\mu\bf k}(x,t)}{\partial t}\right|_{\rm
    dif}\\\mbox{}&&+\left.\frac{\partial\rho_{\mu\bf k}(x,t)}{\partial t}\right|_{\rm
    coh}+\left.\frac{\partial\rho_{\mu\bf k}(x,t)}{\partial t}\right|_{\rm
    scat}.
\label{ksbee}
\end{eqnarray}
Here $\rho_{\mu\bf k}(x,t)$ represent the density matrices of
electrons with relative momentum ${\bf k}$ in valley ${\mu}$ at
position $x$ and time $t$. $\left.\frac{\partial\rho_{\mu\bf k}(x,t)}{\partial t}\right|_{\rm
  dri}=\frac{eE}{\hbar}\frac{\partial \rho_{\mu\bf k}(x,t)}{\partial
  k_x}$ are the driving terms from the external electric field (the
fluctuation of electron density is neglected and thus the total
electric field is taken to be the external one). The diffusion terms
due to the spatial gradient are
\begin{eqnarray}\nonumber
 \left.\frac{\partial\rho_{\mu\bf k}(x,t)}{\partial t}\right|_{\rm
  dif}&=&-\frac{\partial \varepsilon_{\bf k}}{\hbar\partial k_x}\frac{\partial \rho_{{\mu\bf k}}(x,t)}{\partial
  x}\\
&=&-v_{\rm F}\cos\theta_{\bf k}\frac{\partial \rho_{{\mu\bf
       k}}(x,t)}{\partial x}.
\end{eqnarray}
 $\left.\frac{\partial\rho_{\mu\bf k}(x,t)}{\partial t}\right|_{\rm
  coh}$ and $\left.\frac{\partial\rho_{\mu\bf k}(x,t)}{\partial t}\right|_{\rm
  scat}$ are the coherent and scattering terms, respectively. Their
expressions can be found in Ref.~\onlinecite{yzhou}. In the 
steady-state scattering-free spin diffusion, the spatial spin precession
frequency, given by Eq.~(\ref{frequency}), is immediately obtained
according to the KSBEs.\cite{chengtrans1,chengtrans2,zhangtrans}

\section{Spin relaxation and spin diffusion and transport} 
In the following, we first study the spin relaxation in graphene by
fitting the experimental data from Pi {\sl et
  al.}\cite{Pi} to obtain information on impurities (including the effective density as
well as the distance from the graphene layer) and the chemical
  doping dependence of the Rashba spin-orbit coupling coefficient. 
We then use the information to study the spin diffusion and transport
in graphene, first analytically for the case with strong
electron-impurity scattering only, and then numerically with {\em all} the scattering explicitly included. 

\subsection{Spin relaxation time}
We fit the chemical doping dependence of spin relaxation time and diffusion coefficient from Pi {\sl et al.} [Fig.~3(c)
in Ref.~\onlinecite{Pi}] to establish: (i) the density and typical distance from the
graphene layer of charged impurities initially present in the substrate and those of the
chemical doping adatoms; and (ii) the dependence of $\alpha_{\rm R}$
on chemical doping. The electron density $N_e$ is $2.9\times 10^{12}$~cm$^{-2}$
and the temperature $T$ is 18~K.\cite{Pi} The electrons are 
initially polarized in the $x$-$y$ plane\cite{Pi} with the
polarization $P_0$ assumed to be 0.05. To perform the fitting, the KSBEs are solved in the time domain under spatial uniform
case, as carried out recently by Zhou and Wu in the ultraclean
graphene monolayer.\cite{yzhou} (An analytical study of spin
relaxation time in graphene is also given in Appendix~\ref{appa}.) The diffusion coefficient $D$ given by
Pi {\sl et al.} is actually for spin instead of charge, although it is treated as the charge diffusion
coefficient in the experiment.\cite{Pi} In fact, these two coefficients are usually close to each
other and J\'ozsa {\sl et al.} found this most likely to be the case in 
graphene when the electron density is high ($\sim 3\times
10^{12}$~cm$^{-2}$)\cite{Jozsa_09} due to the weak electron-electron
Coulomb scattering.\cite{yzhou} Therefore we fit the experimental
data with the charge diffusion coefficient $D=\frac{\sqrt{\pi N_e}}{2e}\hbar v_{\rm F}\mu_e$, where $\mu_e$ is the electron 
mobility obtained under a small in-plane electric field.\cite{yzhou}

\begin{figure}[htb]
  {\includegraphics[width=8cm]{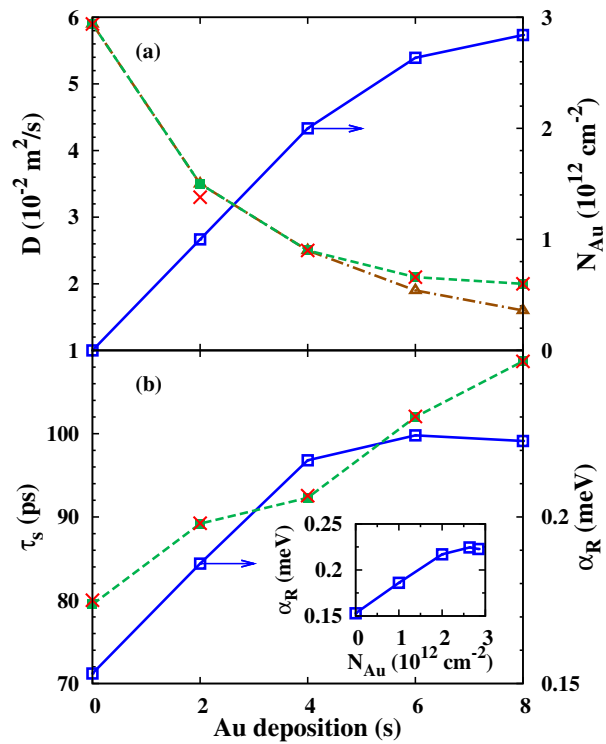}}
  
  \caption{(Color online) (a) Chain curve with triangles: deposition time dependence of
    calculated diffusion coefficient, with the 
    Au density growing linearly with the deposition time with a fixed rate
    of 5$\times 10^{11}$ atom/(cm$^2\cdot$s).\cite{Pi}  Dashed curve
    with closed squares: deposition time dependence of calculated diffusion coefficient,
    with the deposition time dependence of Au density given by
 the solid curve (the scale is on the right-hand side of
    the frame). Crosses: experimental data from Pi {\sl et
      al.}.\cite{Pi} (b) Dashed curve with closed squares:  deposition time dependence of
    calculated spin relaxation time, with the
    deposition time dependence of $\alpha_{\rm R}$ shown by the
    solid curve with open squares (the scale is on the right-hand side of the
    frame). Crosses: experimental data from Pi {\sl et
      al.}.\cite{Pi} Inset of (b): dependence of $\alpha_{\rm R}$ on Au density.}
  \label{figzw1}
\end{figure}

We first make use of the value of $D\approx 0.059$~m$^2$/s for the case without surface
chemical doping given in Ref.~\onlinecite{Pi} to explore the
information of impurities initially present in the
substrate. This single value of $D$ is not sufficient for
us to fix both the effective density and distance from the graphene
layer of these impurities. However, these details are not essential
and we just choose two proper parameters, e.g., $N_s=2.1\times
10^{12}$~cm$^{-2}$ and $d_s=0.7$~nm, to ensure $D\approx
0.059$~m$^2$/s. The surface chemical doping deposits Au atoms on the graphene
surface with a growth rate of 5$\times 10^{11}$ atom/(cm$^2\cdot$s).\cite{Pi} By fitting the
deposition time (adatom density) dependence of $D$,\cite{Pi} the
distance of adatoms from the graphene layer $d_{\rm Au}$ is obtained 
to be about 0.2~nm. Nevertheless, the fitting does not confirm with the
experimental data well when the deposition time exceeds 4~s [compare the
fitting data (chain curve with triangles)
to the experimental data (crosses) in
Fig.~\ref{figzw1}(a)]. This indicates that the effective density of
adatoms does not increase linearly with time any more when the doping has been  
performed for several seconds. Therefore, when the doping time is longer than
4~s, we choose the proper density of adatoms to reproduce the 
experimental diffusion coefficient. In Fig.~\ref{figzw1}(a), the deposition time dependence of
Au density is plotted by the solid curve with open squares (the scale is on
the right-hand side of the frame) and that of the calculated diffusion
coefficient is shown by the dashed curve with closed squares. 

With the parameters for two kinds of impurities obtained, we
then fit the spin relaxation time $\tau_s$ to obtain $\alpha_{\rm R}$ under different deposition times. In
Fig.~\ref{figzw1}(b), the deposition time dependence of fitted $\alpha_{\rm R}$ is
shown by the solid curve with open squares (the scale is on the
right-hand side of the frame) and that of the calculated spin
relaxation time is shown by the dashed curve with closed squares. The
crosses represent the experimental spin relaxation times under
different deposition times. In the
inset of Fig.~\ref{figzw1}(b), we also plot the dependence of
$\alpha_{\rm R}$ on Au density $N_{\rm Au}$. It is shown that $\alpha_{\rm R}$
increases approximately linearly with Au density when
the latter is not so high. The fitted value of $\alpha_{\rm R}$ is comparable to the
value estimated by Ertler {\sl et al.} when taking account of the adatoms, i.e., 0.3~meV.\cite{Fabian_SR} It is
noted that $\alpha_{\rm R}\tau_p/\hbar$ has the largest value
0.027$\ll$1 with $\tau_p=\sqrt{N_e\pi}\frac{\hbar}{ev_{\rm
    F}}\mu_e=0.12$~ps (Ref.~\onlinecite{tan}) when $N_{\rm
Au}=0$. Therefore, the electron system is in the strong
scattering limit (the electron-impurity scattering is dominant), let alone when the temperature is increased or the
chemical doping is performed. It is necessary to point out that in the experiment the gate voltage $V_g$ is adjusted to keep
$N_e$ constant during chemical doping as adatoms also donate
electrons to the graphene layer.\cite{Pi} However, $V_g$ does not exceed 200~V and the term $\zeta
E_z=\zeta V_g/a$ is at least two orders of magnitude smaller than $\alpha_{\rm R}$. Therefore $\alpha_{\rm R}
\approx\eta$ and is solely determined by the impurities. When $N_{\rm
  Au}=0$, $\eta=0.153$~meV and is contributed by the impurities in the 
substrate. 

\subsection{Spin diffusion and transport: analytical study}
\subsubsection{Spin diffusion}
\label{sdanalytical}
In this section we study the spin diffusion in
graphene analytically for the case with only the electron-impurity 
scattering. No external electric field is
present. We first perform the Fourier
transformation of the steady-state KSBEs with respect to the polar angle $\theta_{\bf k}$
and then retain the equations involving the lowest three orders.\cite{zhangtrans} The neglect of the higher orders will
not lose much information in the strong scattering limit where the
electron distribution approaches isotropy in the momentum space. As a result the
following second-order differential equation about $\rho_{\mu k}^0(x)$
[$\rho_{\mu k}^l(x)=\frac{1}{2\pi}\int_0^{2\pi}d\theta_{\bf k}
\rho_{\mu{\bf
  k}}(x)e^{-il\theta_{\bf k}}$ and $\rho_{\mu {\bf
  k}}(x)\equiv\rho_{\mu {\bf k}}(x,+\infty)$] is obtained:
\begin{eqnarray}\nonumber
\hspace{-0.3 cm}&&\partial^2_x\rho_{\mu k}^0(x)+i\frac{2\alpha_{\rm R}}{\hbar v_{\rm F}}[\sigma_y,\partial_x\rho_{\mu
  k}^0(x)]-\frac{\alpha_{\rm R}^2}{\hbar^2v_{\rm F}^2}[\sigma_x,[\sigma_x,\rho_{\mu
  k}^0(x)]]\\ 
\hspace{-0.3 cm}&&\mbox{}-\frac{\alpha_{\rm R}^2}{\hbar^2v_{\rm
      F}^2}[\sigma_y,[\sigma_y,\rho_{\mu k}^0(x)]]=0.
\label{dif}
\end{eqnarray}
It is noted that with only the lowest three orders of $\rho_{\mu
  k}^l(x)$ considered from the beginning, the electron-impurity 
scattering time is actually absent from the above equation (refer to
  Appendix~\ref{appb} for detail). This indicates that in the strong scattering limit the spin diffusion 
becomes insensitive to scattering in graphene. We define the ``spin vector'' as ${\bf
  S}_{\mu k}^0(x)=\mbox{Tr}[\rho_k^0(x){\bgreek \sigma}]$ and ${\bf S}_{\mu k}^0(x)$ can be solved from
Eq.~(\ref{dif}) with boundary conditions (refer to
  Appendix~\ref{appb} for detail). Then one can calculate the
total spin signal contributed by all the different electron states in two valleys
as \begin{eqnarray}\nonumber
{\bf S}(x)&=&\frac{1}{4\pi^2}\sum_\mu\int_0^{+\infty}dk\int_{0}^{2\pi}d\theta_{\bf
  k}k\mbox{Tr}[\rho_{\bf k}(x){\bgreek \sigma}]\\
&=&\frac{1}{\pi}\int_0^{+\infty}dkk{\bf S}_{\mu k}^0(x).
\label{sum}
\end{eqnarray}
In the following we present the solutions of ${\bf S}(x)$ under three typical boundary
conditions. 

For boundary condition (I) ${\bf S}_{\mu k}^0(0)=({S_{\mu k}^0}(0),
0, 0)$ and  ${\bf S}_{\mu  k}^0(+\infty)=0$, which corresponds to the case with the injected
spins polarized along the $x$-axis, 
\begin{eqnarray}
{\bf S}(x)=S(0)e^{-x/l_x}\left(\begin{array}{c}\sqrt{1+\Delta^2}\sin(\omega x+\phi) \\0\\ c_1\sin(\omega x)\end{array}\right).
\label{sx}
\end{eqnarray}
For boundary condition (II) ${\bf S}_{\mu k}^0(0)=(0, {S_{\mu k}^0}(0),
0)$ and  ${\bf S}_{\mu  k}^0(+\infty)=0$,  
\begin{eqnarray}
 {\bf S}(x)=S(0)e^{-x/l_y}\left(\begin{array}{c}0 \\1\\ 0\end{array}\right)
 \label{sy}.
\end{eqnarray}
For boundary condition (III) ${\bf S}_{\mu k}^0(0)=(0, 0, {S_{\mu
    k}^0}(0))$ and  ${\bf S}_{\mu  k}^0(+\infty)=0$, 
\begin{eqnarray}
    {\bf S}(x)=S(0)e^{-x/l_z}\left(\begin{array}{c}c_2\sin(\omega
  x)\\0\\ -\sqrt{1+\Delta^2}\sin(\omega x-\phi)\end{array}\right).
    \label{sz}
\end{eqnarray}
In the above equations 
\begin{eqnarray}
 S(0)=\frac{1}{\pi}\int_0^{+\infty}dkkS_{\mu
    k}^0(0)
\label{s0}
\end{eqnarray}
and 
\begin{eqnarray}
&& l_x=l_z=\frac{\sqrt{7}}{(2\sqrt{2}-1)\sqrt{1+2\sqrt{2}}}\frac{\hbar
  v_{\rm F}}{\alpha_{\rm R}},
\label{lx}\\
&& l_y=\frac{\hbar v_{\rm F}}{2\alpha_{\rm R}},
\label{ly}\\
&& \omega=\sqrt{1+2\sqrt{2}}\frac{\alpha_{\rm R}
}{\hbar v_{\rm F}}.
\end{eqnarray}
$c_1=-\frac{4}{(1+\sqrt{2})\sqrt{1+2\sqrt{2}}}$, $c_2=\frac{(20\sqrt{2}-24)\sqrt{1+2\sqrt{2}}}{7}$,
$\Delta=\frac{8\sqrt{2}-11}{\sqrt{7}}$ and
$\phi=\arctan\frac{1}{\Delta}$. It is noted that the spin precession
frequency given by the simplified model is $\omega\approx
1.96\frac{\alpha_{\rm R}}{\hbar v_{\rm F}}$, a little smaller than $|\langle {\bgreek \omega}_{\bf
  k}\rangle|=\frac{2\alpha_{\rm R}}{\hbar v_{\rm F}}$ due to the
approximations made here.

From Eqs.~(\ref{sx})-(\ref{sz}) one notices that in the strong scattering limit, the
spin diffusion is not only insensitive to the scattering, but also unrelated to temperature $T$ and electron
density $N_e$. Nevertheless, the coefficient $\alpha_{\rm R}$ may
depend on $T$ and/or $N_e$, with the relation unclear so far. For
simplicity we assume $\alpha_{\rm R}$ to be independent of $T$ and
$N_e$ in this work. As a result, the spin diffusion in the strong 
scattering limit is uniquely determined by $\alpha_{\rm
  R}$, which is only modulated by chemical
doping. Eqs.~(\ref{sx})-(\ref{sz}) indicate a strong anisotropy of spin
diffusion with respect to the spin-polarization direction. For the cases with the injected spins polarized along the  
$x$- and $z$-axis, both the spin signals show an
exponential decay in the magnitude accompanying with the precession in the $x$-$z$ plane. The spin precessions have the same frequency
$\omega$ except for a phase difference. However, when the 
injected spins are polarized along the $y$-axis, the spin
signal decays exponentially without any precession, i.e., it is
  bound along the $y$-axis. The above phenomena are
  understood by noticing that 
  the mean effective magnetic field felt by the diffusing electrons is
  along the $y$-axis as $\langle {\bgreek \omega}_{\bf
    k}\rangle=\frac{2\alpha_{\rm R}}{\hbar v_{\rm F}}(0, 1, 0)$. In the non-local spin valve
experiments, the spin diffusion length is usually measured from the
exponential decay of spin signal with the increasing spacing between the central
spin-injector and -detector ferromagnetic electrodes.\cite{Popinciuc,han} In
these experiments, the ferromagnetic electrodes happen to be
magnetized along the $y$-axis and therefore the injected and
detected spin polarizations are both along the $y$-axis. With
such configuration, the exponential decay of spin signal with
increasing spacing between the electrodes can be well observed. However, if the
injected spins are polarized in the $x$-$z$ plane, the spatial spin precession is expected to be
detected. 

Besides the anisotropy of spin precession,  the spin diffusion length also shows an
anisotropy as 
\begin{equation}
l_x=l_z\approx 1.48 l_y
\end{equation}
with
  $l_y=\hbar v_{\rm   F}/(2\alpha_{\rm R})$.
In fact, when the injected spins are polarized along any other
  direction in the $x$-$z$ plane, the spin
  diffusion length is all the same as $l_x$ ($l_z$)
 [for this case the solution of
  ${\bf S}(x)$ is the combination of Eqs.~(\ref{sx}) and (\ref{sz})]. 
However, based on the widely utilized two-component drift-diffusion
model\cite{dd,yu,fabiandd,fabiandd1,fabiandd2} which gives $l_s=\sqrt{D\tau_s}$
 [Eq.~(\ref{d6}) in Appendix~\ref{dde}], one 
may expect that the spin diffusion lengths satisfy
\begin{equation}
l_x=l_y=\sqrt{2}l_z=\hbar v_{\rm F}/(2\alpha_{\rm R})
\end{equation}
as the spin relaxation times in time domain follow (refer to
Appendix~\ref{appa})
\begin{equation}
\tau_x=\tau_y=2\tau_z=\hbar^2/(2\alpha_{\rm R}^2\tau_p)
\end{equation}
and $D=v_{\rm F}^2\tau_p/2$. It is noted that only when the
  injected spins are polarized along the $y$-axis, for which
no spin precession exists, the two-component drift-diffusion model gives the result in consistence with that from the
KSBEs, i.e., 
\begin{equation}
l_y=\hbar v_{\rm F}/(2\alpha_{\rm R}).
\label{consistence}
\end{equation}
The discrepancy in the anisotropies given by the KSBEs and the
two-component drift-diffusion model strongly
indicates the inadequacy of the two-component drift-diffusion model. Due to the loss
of the off-diagonal spin components, i.e., the spin
coherence, the two-component drift-diffusion model not only fails to predict the
spin precession in spatial domain in the absence of an external magnetic
field, but also incorrectly inherits the anisotropy from the spin
relaxation in time domain. We emphasize that the reason for the
different anisotropic properties of spin diffusion in spatial domain and spin relaxation in
time domain is that the inhomogeneous broadening is quite different in
these two cases. In spatial domain the inhomogeneous broadening
governing the spin diffusion arises from the ${\bf k}$ dependence of
${\bgreek \omega}_{\bf k}$, while in time domain from that of
${\bgreek \Omega}_{\bf k}$. Popinciuc {\sl et al.} reported the 
relationship between the in-plane and out-of-plane spin relaxation 
times directly from the anisotropy of spin diffusion via the
two-component drift-diffusion model.\cite{Popinciuc} However, based on the above discussion,
one may realize that studying the anisotropy of spin relaxation time
in such a way can be incorrect. Finally, from another point
  of view, if the two-component drift-diffusion model is still used, then in order to reflect the
  correct anisotropy of spin diffusion, the spin diffusion
  coefficient has to differ from the charge diffusion
  coefficient and shows an anisotropy as $D_x=0.5D_z\approx 2.2 D_y$
  with $D_y=v_{\rm F}^2\tau_p/2$.

It should be pointed out that all the above analysis and conclusion also
apply to the electron system where the energy
spectrum is parabolic in momentum and the linear Rashba spin-orbit coupling term
$\bm{\Omega}_{\bf  k}\propto k(-\sin\theta_{\bf k},
\cos\theta_{\bf k},0)$ is dominant, such as that in the asymmetric InAs
quantum wells.\cite{prepare} That is because the steady-state scattering-free
 spatial spin
precession frequency ${\bgreek \omega}_{\bf k}$ in this system has the similar momentum
dependence as shown in Eq.~(\ref{frequency}).\cite{prepare} However, for electron system in
the absence of the DP term but under a magnetic field perpendicular
to both the spin polarization and spin transport directions such as in bulk
silicon\cite{appelbaum} and symmetric silicon quantum
wells,\cite{zhangtrans} or with the  Dresselhaus
term\cite{dresselhaus} containing the cubic dependence on momentum such as in GaAs quantum
wells,\cite{chengtrans1,chengtrans2} the situation is quite
different as ${\bgreek \omega}_{\bf k}$ depends on the magnitude of
momentum. In fact, it has been revealed in the symmetric silicon
quantum wells under an in-plane magnetic field that the scattering can
suppress spin diffusion effectively in the strong scattering limit.\cite{zhangtrans}

\subsubsection{Spin transport}
\label{stanalytical}
We further take account of the electric field along the $x$-axis to study the spin transport. Still only the strong electron-impurity 
scattering is included.  The second-order differential equation about
$\rho_{\mu k}^0(x)$, corresponding to Eq.~(\ref{dif}) but including
the driving term, reads (refer to Appendix~\ref{appc})

\begin{eqnarray}\nonumber
\hspace{-0.3 cm}&&\partial^2_x\rho_{\mu k}^0(x)+i\frac{2\alpha_{\rm R}}{\hbar v_{\rm F}}[\sigma_y,\partial_x\rho_{\mu
  k}^0(x)]-\frac{\alpha_{\rm R}^2}{\hbar^2v_{\rm F}^2}[\sigma_x,[\sigma_x,\rho_{\mu
  k}^0(x)]]\\ \nonumber \hspace{-0.3 cm}&&\mbox{}-\frac{\alpha_{\rm R}^2}{\hbar^2v_{\rm
      F}^2}[\sigma_y,[\sigma_y,\rho_{\mu
        k}^0(x)]]-eE\partial_x\partial_{\varepsilon_{\bf k}}\rho_{\mu
    k}^0(x)\\
\hspace{-0.3 cm}&&\mbox{}-i\frac{\alpha_{\rm R}eE}{\hbar v_{\rm  F}}[\sigma_y,\partial_{\varepsilon_{\bf k}}\rho_{\mu
    k}^0(x)]=0.
\label{transport}
\end{eqnarray}
It should be pointed out that when the electric field is so large that
the electron density matrices $\rho_{\mu{\bf k}}(x,+\infty)$ become
strongly anisotropic due to the driving of the electric field,
retaining only the lowest three orders of $\rho_{\mu
  k}^l(x)$ to obtain the above equation of $\rho_{\mu k}^0(x)$ may not
be sufficient. The second-order differential equation about ${\bf S}_{\mu k}^0(x)$ is
obtained from the above equation and that about ${\bf S}(x)$ can be 
obtained by further summing over ${\bf k}$ and $\mu$ (refer to
Appendix~\ref{appc}). With the same three different typical boundary conditions presented in the
previous section, ${\bf S}(x)$ is solved to have the same form as
Eqs.~(\ref{sx})-(\ref{sz}) except that the parameters are now electric-field dependent. Explicitly,
\begin{eqnarray}
\hspace{-1 cm}&&
  l_x^\prime=l_z^\prime=\frac{1}{{\mathcal E}/2+F({\mathcal E})}\frac{\hbar v_{\rm F}}{\alpha_{\rm  R}},
\label{lxtrans}\\
\hspace{-1 cm}&& l_y^\prime=\frac{1}{{\mathcal E}/2+\sqrt{4+{\mathcal E}^2/4}}\frac{\hbar v_{\rm F}}{\alpha_{\rm  R}},
\label{lytrans}\\
\hspace{-1 cm}&& \omega^\prime=G({\mathcal E})\frac{\alpha_{\rm R}}{\hbar v_{\rm F}},
\end{eqnarray}
$c_1^\prime=-\frac{1}{2}\frac{\sqrt{{\mathcal E}^4+48{\mathcal E}^2+512}}{\sqrt{{\mathcal E}^2+7}F({\mathcal E})+5G({\mathcal E})}$,
$c_2^\prime=\frac{1}{2}\frac{\sqrt{{\mathcal E}^4+48{\mathcal E}^2+512}}{\sqrt{{\mathcal E}^2+7}F({\mathcal E})+3G({\mathcal E})}$,
$\Delta^\prime=\frac{5F({\mathcal E})-\sqrt{{\mathcal E}^2+7}G({\mathcal E})}{\sqrt{{\mathcal E}^2+7}F({\mathcal E})+5G({\mathcal E})}$
and $\phi^\prime=\arctan\frac{1}{\Delta^\prime}$. In the above equations, 
\begin{eqnarray}
{\mathcal E}&=&\frac{eE}{S(0)\pi\alpha_{\rm R}\beta\hbar v_{\rm
    F}}\ln\frac{1+e^{\beta\mu_{\uparrow}}}{1+e^{\beta\mu_{\downarrow}}},
\label{chi}\\\nonumber
F({\mathcal E})&=&\frac{\sqrt{{\mathcal E}^4+48{\mathcal E}^2+512}+{\mathcal E}^2-8}{16\sqrt{2}\sqrt{{\mathcal E}^2+7}}\\\mbox{}&&\times\sqrt{\sqrt{{\mathcal E}^4+48{\mathcal E}^2+512}-{\mathcal E}^2+8},\\
G({\mathcal E})&=&\sqrt{1-{\mathcal E}^2/8+\sqrt{{\mathcal E}^4+48{\mathcal E}^2+512}/8}.
\end{eqnarray}
Here $\beta=1/(k_BT)$ and $\mu_\uparrow$ ($\mu_\downarrow$) is the
chemical potential of electrons with spin parallel (antiparallel) to
the spin-polarization direction. It is noted that when the electric
field is absent, i.e., ${\mathcal E}=0$, all the above solutions recover those
presented in the previous section.

In most conditions (such as in the present work) electrons in graphene are highly degenerate. In the degenerate
limit with small spin polarization, ${\mathcal E}\approx \frac{eE}{\alpha_{\rm
    R}k_{\rm F}}$, where $k_{\rm F}=\sqrt{\pi N_e}$ is the magnitude
of the Fermi momentum of unpolarized electrons with density being $N_e$
(Appendix~\ref{appc}). Differing from the spin diffusion without
electric field, the spin transport becomes sensitive to electron
density as ${\mathcal E}$ depends on the electron density. In the
nondegenerate limit, ${\mathcal E}\approx \frac{eE\beta\hbar v_{\rm 
    F}}{\alpha_{\rm R}}$ (Appendix~\ref{appc}) and the
spin transport becomes sensitive to temperature rather than electron
density. Moreover, with this value of ${\mathcal E}$, Eq.~(\ref{lytrans})
becomes
\begin{equation}
l_y^\prime=\Big[eE\beta/2+\sqrt{e^2E^2\beta^2/4+1/l_y^2}\Big]^{-1},
\label{ddly}
\end{equation}
where $l_y$ is the spin diffusion length without electric field
[Eq.~({\ref{ly}})]. This result recovers that from the two-component drift-diffusion
model, which apparently fails to correctly reflect the anisotropy of spin
transport.\cite{yu,jozsa_08} Therefore, our investigation again indicates that
only when the spatial spin precession is absent, the two-component drift-diffusion model
gives the appropriate depiction of spin transport.

\begin{figure}[ht]
   \hspace{-0.5cm} {\includegraphics[width=9cm]{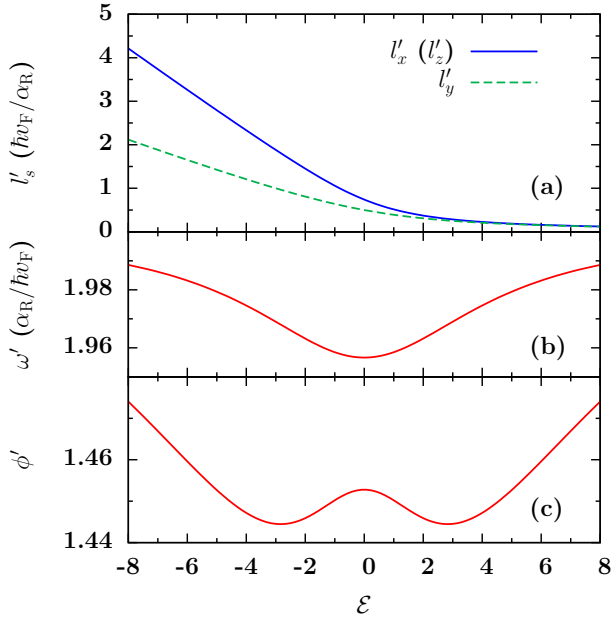}}
    \caption{(Color online) The dependence of (a) spin transport length
    $l_{x,y,z}^\prime$, (b) spin precession frequency $\omega^\prime$
     and (c) phase angle $\phi^\prime$ on ${\mathcal E}$.}
  \label{figzw2}
\end{figure}

In Fig.~\ref{figzw2} we plot the dependence of
$l_{x,y,z}^\prime$, $\omega^\prime$ and $\phi^\prime$ on
${\mathcal E}$. From Fig.~\ref{figzw2}(a), one notices that the spin transport
length decreases with increasing ${\mathcal E}$ (${\mathcal E}\approx\frac{eE}{\alpha_{\rm
    R}k_{\rm F}}$). On one hand, this means that when the
electron density is fixed (e.g., $N_e=10^{12}$~cm$^{-2}$, for which
the variation of ${\mathcal E}$ from $-8$ to 8 corresponds to a variation of
$E$ from about $-2.2$ to 2.2~kV/cm), the spin
transport is suppressed (enhanced) by increasing the electric field parallel (antiparallel) to the spin injection
direction. On the other hand, this also means that when the non-zero electric
field parallel (antiparallel) to the spin injection
direction is fixed, the spin transport is enhanced (suppressed) by
increasing electron density. Fig.~\ref{figzw2}(b) and (c) indicate that the spin precession
frequency $\omega^\prime$ and the phase angle $\phi^\prime$ vary with
${\mathcal E}$ marginally (with a variation $\sim$
2~\%). In fact, when $|{\mathcal E}|$ becomes even larger, both 
$\omega^\prime$ and $\phi^\prime$ quickly saturate ($\omega^\prime$
approaches $\frac{2\alpha_{\rm R}}{\hbar v_{\rm F}}$ and $\phi^\prime$
approaches $\pi/2$). Therefore, the spin precession pattern in spatial
domain is insensitive to the electric field or the electron density. 

\subsection{Spin diffusion and transport: numerical results}
The KSBEs need to be solved numerically in order to take full account
of all the different kinds of scattering as well as the large electric
field. To numerically solve the KSBEs, the initial conditions are set as
\begin{eqnarray}
&&\rho_{\mu \bf k}(0,0)=\frac{F^0_{{\bf k}\uparrow}+F^0_{{\bf
      k}\downarrow}}{2}+\frac{F^0_{{\bf k}\uparrow}-F^0_{{\bf
      k}\downarrow}}{2}{\bf\hat n}\cdot{\bgreek \sigma},\\
&&\rho_{\mu \bf k}(x>0,0)=\frac{F^L_{{\bf k}\uparrow}+F^L_{{\bf
      k}\downarrow}}{2},\\
&&\sum_{\mu\bf  k}\mbox{Tr}[\rho_{\mu\bf k}(0,0){\bf \hat{n}}\cdot{\bgreek \sigma}]/\sum_{\mu\bf
  k}\mbox{Tr}[\rho_{\mu\bf k}(0,0)]=P_0,
\end{eqnarray}
and the two-side injection boundary conditions\cite{chengtrans1,chengtrans2} are
\begin{eqnarray}
\rho_{\mu \bf k}(0,t)|_{k_x>0}&=&\frac{F^0_{{\bf k}\uparrow}+F^0_{{\bf
      k}\downarrow}}{2}+\frac{F^0_{{\bf k}\uparrow}-F^0_{{\bf
      k}\downarrow}}{2}{\bf\hat n}\cdot{\bgreek \sigma},\\
\rho_{\mu \bf k}(L,t)|_{k_x<0}&=&\frac{F^L_{{\bf k}\uparrow}+F^L_{{\bf
      k}\downarrow}}{2}.
\end{eqnarray}
Here the injected spins at left boundary $x=0$ are assumed to be polarized along ${\bf\hat n}$ with
polarization $P_0=0.05$. $x=L$ stands for the right boundary with $L$ much longer than the
spin diffusion or transport length. 
$F^{0,L}_{{\bf k}\uparrow,\downarrow}$ are the
 Fermi distributions of electrons at the
two boundaries when the external electric field is absent. When the
electric field is present, $F^{0,L}_{{\bf k}\uparrow,\downarrow}$ then 
stand for the drifted Fermi distributions of hot
electrons.\cite{yzhou} In the previous analytical study the boundary conditions are in fact approximated as
  the single-side injection case. This approximation works well when the
  scattering is strong.\cite{chengtrans1} By numerically solving the
KSBEs, the steady-state distribution of spin polarization along
${\bf\hat n}$ is obtained as $P(x)=\sum_{\mu\bf
  k}\mbox{Tr}[\rho_{\mu\bf k}(x,+\infty){\bf \hat{n}}\cdot{\bgreek
    \sigma}]/\sum_{\mu\bf k}\mbox{Tr}[\rho_{\mu\bf k}(x,+\infty)]$ and
  then the spin diffusion or transport length is determined from the exponential 
decay of $P(x)$ (or its envelope) along the $x$-axis. 

\begin{figure}[ht]
    {\includegraphics[width=8cm]{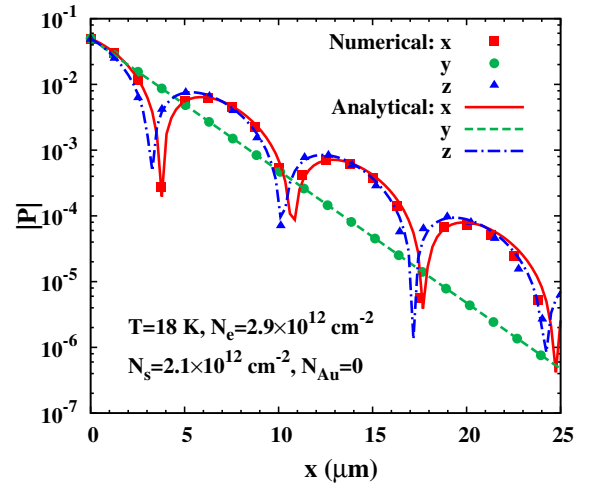}}
    \caption{(Color online) The absolute value of steady-state spin polarization
      $|P|$ versus position $x$ with the injected spins polarized
      along the $x$-, $y$- and $z$-axis,
      respectively. The squares, circles and triangles are obtained by numerically solving the KSBEs with
      $T=18$~K, $N_e=2.9\times 10^{12}$~cm$^{-2}$, $N_s=2.1\times
      10^{12}$~cm$^{-2}$ and $N_{\rm Au}=0$. The solid, dashed and chain curves are
      calculated from Eqs.~(\ref{sx})-(\ref{sz}) with $P(x)=S_x(x)/N_e$, $S_y(x)/N_e$ and $S_z(x)/N_e$, respectively. }
  \label{figzw3}
\end{figure} 

\subsubsection{Anisotropic spin diffusion}
\label{cases}

As revealed by the analytical model, the spin diffusion shows
anisotropic properties with respect to the polarization direction of
injected spins. In
Fig.~\ref{figzw3}, we show the spatial distribution of the absolute
value of the steady-state spin polarization
$|P|$ for the cases with the injected spins polarized 
along the $x$-, $y$- and $z$-axis,
respectively. $N_{\rm Au}=0$, with which $\alpha_{\rm R}=0.153$~meV. 
The squares, circles and triangles are obtained by numerically solving
the KSBEs while the solid, dashed and chain curves are calculated by
Eqs.~(\ref{sx})-(\ref{sz}). When $\alpha_{\rm R}=0.153$~meV the analytical model gives $l_x=l_z\approx 3.18$~$\mu$m,
$l_y\approx 2.16$~$\mu$m and $\omega\approx 0.45$~$\mu$m$^{-1}$. The
anisotropy of spin diffusion is clearly shown in this figure. It is noted that the simplified analytical model almost
perfectly recovers the numerical results [except that the spin
    precession frequencies for both cases with the injected spins polarized 
along the $x$- and $z$-axis are numerically shown
to be closer to $2\frac{\alpha_{\rm R}}{\hbar v_{\rm F}}$ rather
than $1.96\frac{\alpha_{\rm R}}{\hbar v_{\rm F}}$ given by the
analytical study (the difference is expected from the approximations
made in the analytical analysis)]. In fact, further numerical calculations show that varying $T$ from 18 to 300~K
and/or $N_e$ from 0.5 to $2.9\times 10^{12}$~cm$^{-2}$ changes the
numerical results marginally. This is consistent with the conclusion from 
the analytical model, i.e., the spin diffusion of electrons in graphene is
{\em insensitive} to $T$ and $N_e$ in the strong scattering
limit. As a result, in the strong scattering limit, one can
depict the spin diffusion quite well with the {\em single}
parameter $\alpha_{\rm R}$ via Eqs.~(\ref{sx})-(\ref{sz}). 

\subsubsection{Chemical doping dependence of spin diffusion}

\begin{figure}[ht]
  {\includegraphics[width=8.5cm]{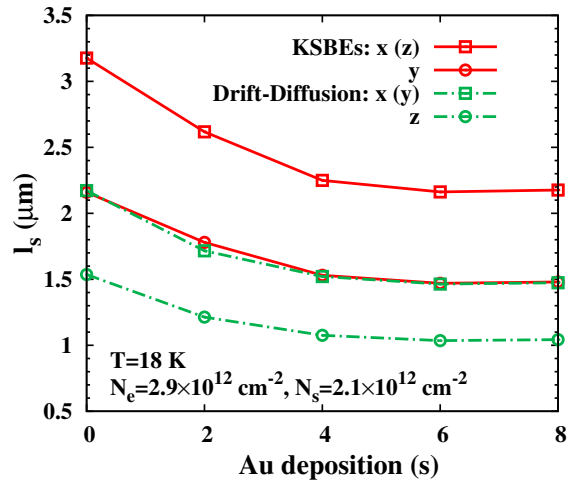}}
  
  \caption{(Color online) Deposition time
dependence of spin diffusion length with the injected spins polarized
along the $x$-, $y$- and $z$-axis,
respectively. The results from the KSBEs (solid curves) and the
two-component drift-diffusion model (chain curves) are both plotted for comparison.}
  \label{figzw4}
\end{figure}

In Fig.~\ref{figzw4}, we plot the deposition time
dependence of spin diffusion length with ${\bf\hat n}={\bf\hat x}$, ${\bf\hat y}$ and ${\bf\hat z}$
respectively by the solid curves. The spin diffusion lengths are
directly obtained from Eqs.~(\ref{lx})-(\ref{ly}). It is shown
  that with the increase of chemical doping time, $\alpha_{\rm R}$
  increases and the spin diffusion length decreases. For comparison, we
also plot the deposition time dependence of spin diffusion length
given by the two-component drift-diffusion model (chain curves), i.e., 
$l_x=l_y=\sqrt{2}l_z=\sqrt{D\tau_x}$,
with $D$ and $\tau_x$ given in Fig.~\ref{figzw1}. The comparison between these two sets of results shows
  that, only when the injected spins are polarized along the $y$-axis,
  the two-component drift-diffusion model 
yields the same result as that from the KSBEs, just as revealed in the
analytical study [refer to Eq.~(\ref{consistence}) and the discussion there].

\subsubsection{Effect of scattering on spin diffusion}
\begin{figure}[bht]
  {\includegraphics[width=8.5cm]{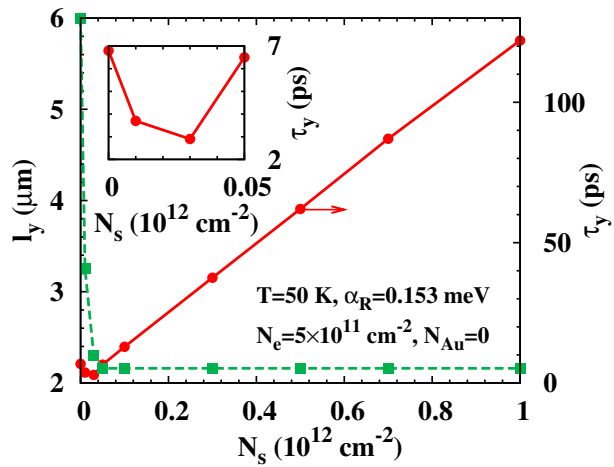}}
  
  \caption{(Color online) The impurity (in the substrate) density dependence of spin
    diffusion length (dashed curve) and spin relaxation time (solid
    curve with the scale on the right-hand side of the frame). The inset shows
    the detail of the solid curve in the small
    density regime. The injected spins are polarized along the
    $y$-axis. $T=50$~K, $N_e=5\times 10^{11}$~cm$^{-2}$ and
    $\alpha_{\rm R}=0.153$~meV.}
  \label{figzw5}
\end{figure}

The electron system under investigation is always in the strong scattering
limit and therefore the spin diffusion becomes insensitive to
scattering. However, the properties of spin diffusion in the weak
scattering limit can be different. In order to investigate the spin
diffusion with scattering strength ranging from the weak to strong
scattering limit, we artificially vary
the impurity density in the substrate from 0 to 10$^{12}$~cm$^{-2}$. 
At the same time, the chemical doping is absent (no adatom)
and $\alpha_{\rm R}$ is kept as a constant, e.g., 0.153~meV. We choose
$T=50$~K, $N_e=5\times 10^{11}$~cm$^{-2}$ and ${\bf\hat n}={\bf\hat
  y}$. In Fig.~\ref{figzw5} we plot the dependence of spin 
diffusion length $l_y$ on the impurity density by the dashed curve. For comparison, we also plot 
the corresponding dependence of spin relaxation time $\tau_y$ on the impurity
density by the solid curve (the scale is on the right-hand side of the frame).
It is seen that with the increase in $N_s$, while $l_y$ decreases obviously in the
weak scattering limit ($N_s\lesssim 0.05\times 10^{12}$~cm$^{-2}$) 
and then saturates in the strong scattering limit, $\tau_y$ first decreases in the
weak scattering limit (refer to the inset for detail) and then 
increases almost linearly in the strong scattering limit.\cite{wu-review} 
The two-component drift-diffusion model is able to capture the dependence of
spin diffusion length on $N_s$ by means of the relation
$l_y=\sqrt{D\tau_y}$: while $D\propto \tau_p\propto 1/N_s$, $\tau_y$
decreases with $N_s$ in the weak scattering limit and $\propto N_s$ in
the strong scattering limit; therefore $l_y$ first decreases with $N_s$
and then becomes insensitive to $N_s$ (the insensitivity of $l_y$ to
$N_s$ in the strong scattering limit is revealed previously by the
analytical study). It should be emphasized that in Fig.~\ref{figzw5}
the results are shown with $\alpha_{\rm R}$ being a constant. In
reality, when one further takes account of the increase of
$\alpha_{\rm R}$ with increasing $N_s$, $l_y$ should always decrease
with increasing $N_s$, from the weak scattering limit to the strong
scattering limit.

 \begin{figure}[ht]
    {\includegraphics[width=8cm]{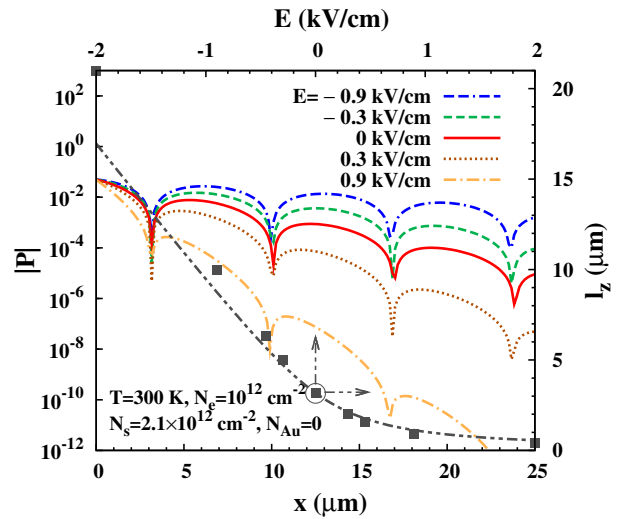}}
    \caption{(Color online) The absolute value of the steady-state spin 
polarization $|P|$ versus position $x$ under different electric fields. 
  The electric field dependence of spin transport length 
      $l_z$ is also plotted with the scale on the right-hand side and top of the
      frame, where  the squares and double-dotted chain curve
are obtained from the numerical calculation and  from Eq.~(\ref{lxtrans}),
respectively. $T=300$~K,
       $N_e=10^{12}$~cm$^{-2}$, $N_s=2.1\times 10^{12}$~cm$^{-2}$ and $N_{\rm Au}=0$.}
   \label{figzw6}
 \end{figure}

 \begin{figure}[htb]
    {\includegraphics[width=8cm]{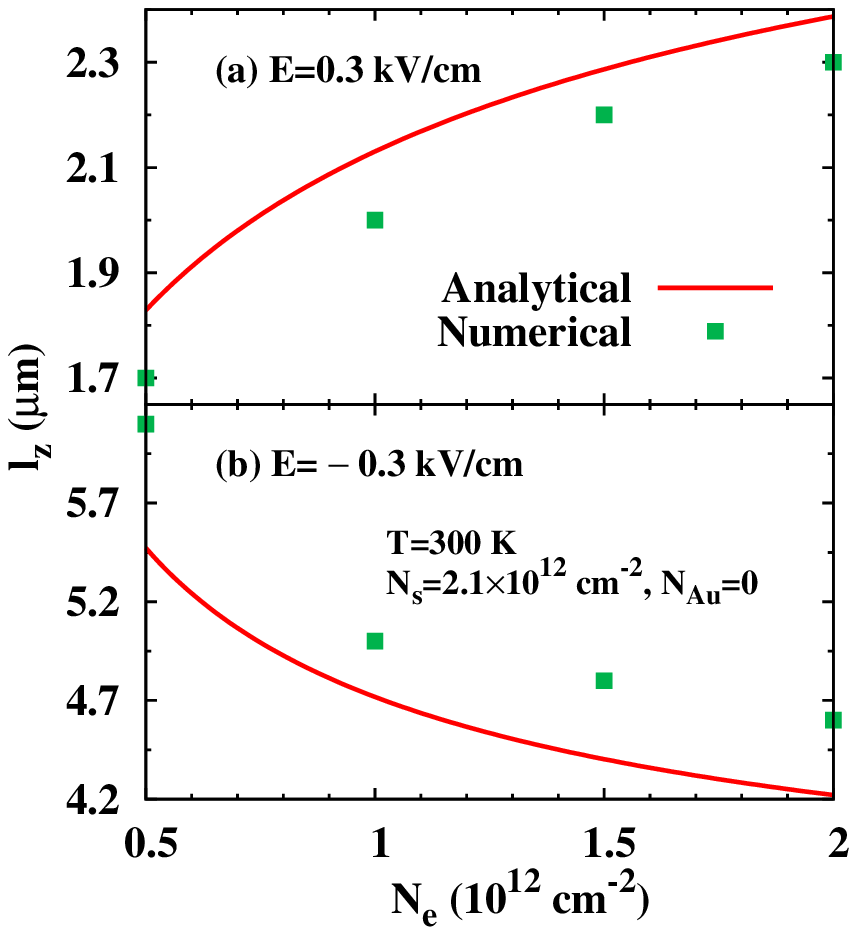}}
    \caption{(Color online)  Electron density dependence of spin
      transport length $l_z$ under electric fields with opposite
      directions: (a) $E=0.3$~kV/cm and (b) $E=-0.3$~kV/cm. The
      squares are from the numerical calculation while the
       curves are from Eq.~(\ref{lxtrans}). $T=300$~K, $N_s=2.1\times 10^{12}$~cm$^{-2}$ and $N_{\rm Au}=0$.}
   \label{figzw7}
 \end{figure} 

\subsubsection {Spin transport under the electric field}

 At last we investigate the spin transport under an electric field along the $x$-axis. $T=300$~K, $N_e=10^{12}$~cm$^{-2}$, $N_s=2.1\times
 10^{12}$~cm$^{-2}$ and $N_{\rm Au}=0$. The injected spins are
 polarized along the $z$-axis. In Fig.~\ref{figzw6} the position dependence of $|P|$ under
 different electric fields as well as the $E$ dependence of $l_z$
 (squares with the scale on the right-hand side and top of the frame) are
 plotted. It is shown that while the spin-precession pattern almost keeps the same with varying
 $E$,  the spin transport length is increased (decreased) by increasing the electric field along the $-x$ ($x$)-direction.\cite{chengtrans1,yu} These results
 are in consistence with the analytical study presented in
 Sec.~\ref{stanalytical}. For comparison, we further plot the $E$
 dependence of $l_z$ from Eq.~(\ref{lxtrans}) by the double-dotted 
chain curve with
 the scale also on the right-hand side and the top of the frame. It is shown that the
 analytical model depicts the spin transport in the low electric-field
 regime well except when the electric field antiparallel to the
 spin-injection direction is large (e.g, a discrepancy reaches 20\% when
 $E$ reaches $-2$~kV/cm).

The electron density dependence of spin
transport is also investigated. In Fig.~\ref{figzw7}, we plot the
density dependence of spin transport length under the electric field parallel
  ($E=0.3$~kV/cm) and antiparallel ($E=-0.3$~kV/cm) to the spin
  transport direction in (a) and (b), respectively. The squares are
  from the numerical calculation and the curves are from Eq.~(\ref{lxtrans}). It is clearly shown that for the cases with
  opposite directions of the electric field, the density dependences
  of spin transport length have opposite tendencies.

\section{Conclusion}

In conclusion, we have investigated the spin diffusion and transport in
graphene monolayer on SiO$_2$ substrate as presented by Pi {\sl et
  al.},\cite{Pi} by means of the KSBE approach.
The substrate (including the impurities initially present) contributes a
Rashba spin-orbit coupling field much stronger than the one
modulated by the electric field perpendicular to the graphene
layer. By surface chemical doping with Au adatoms, the Rashba
spin-orbit coupling coefficient $\alpha_{\rm R}$ is increased. By fitting the
chemical doping dependence of diffusion coefficient and spin
relaxation time,\cite{Pi} we obtain the information on impurities as
well as the chemical doping dependence of $\alpha_{\rm R}$. Our fitting
finds that $\alpha_{\rm R}$ increases linearly from 0.15 to 0.23~meV with increasing Au
density when the latter is not so high. With the necessary parameters obtained from fitting, we
investigate the spin diffusion and transport in graphene both analytically and numerically.

The analytical study with only the electron-impurity scattering
included reveals that in the strong scattering limit (just
as the situation under investigation in the present work), the spin
diffusion is {\em uniquely} determined by $\alpha_{\rm R}$. When the injected spins are polarized along the
$x$-, $y$- and $z$-axis, the spin
diffusion lengths are given by the analytical study with an anisotropy
 as $l_x=l_z\approx0.74\hbar v_{\rm F}/\alpha_{\rm R}$ and $l_y=0.5\hbar v_{\rm
  F}/\alpha_{\rm R}$. Meanwhile, the spatial spin precession
 is present when the injected spins are
polarized in the $x$-$z$ plane 
but absent when the injected spins are
polarized along the $y$-axis. Further numerical calculations
with all the scattering explicitly included show that the
analytical model depicts the spin diffusion pretty well. 

It is noted that the anisotropy of spin diffusion length from the KSBEs
 differs from the one from the two-component drift-diffusion model where
$l_x=l_y=\sqrt{2}l_z=0.5\hbar v_{\rm F}/\alpha_{\rm R}$. 
The qualitative discrepancy indicates the 
inadequacy of the two-component drift-diffusion model due to the neglect of the
off-diagonal spin components, i.e., the spin coherence. In fact, only
when the injected spins are polarized along the $y$-axis and the
spatial spin precession is absent, the two-component drift-diffusion model gives the same
spin diffusion length as the KSBE approach does.

The analytical and numerical study of spin transport under an electric
field parallel or antiparallel to the spin injection direction is also
investigated. In the presence of the electric field, the analytical
  model depicts the spin transport with a small discrepancy which increases with the strength of the electric field.
It is shown that when the electric field is applied, the
spin precession in spatial domain for the cases with the injected spins polarized along the $x$- and $z$-axis remains almost unchanged. However, the spin transport length is increased (decreased)
by increasing the magnitude of the electric field when it is antiparallel (parallel) to the spin transport
direction. Moreover, in the presence of the electric field, the spin
transport becomes sensitive to the electron density, differing from
the case of spin diffusion. The spin transport is enhanced (suppressed) by
increasing electron density when the electric field is parallel
 (antiparallel) to the spin injection direction. 

\begin{acknowledgments}
This work was supported by the National Natural Science
Foundation of China under Grant No. 10725417. One of the authors
(MWW) acknowledges
valuable discussions with J. Fabian and B. J. van Wees.

\end{acknowledgments}

\begin{appendix}
  \section{Spin relaxation in graphene}
  \label{appa}
We consider spin relaxation in graphene under the spatial uniform
case in the absence of the electric field. We only include the electron-impurity scattering. The KSBEs,
Eq.~(\ref{ksbee}), are then simplified to be
\begin{eqnarray}\nonumber
\partial_t\rho_{\mu {\bf k}}(t)=&&-\frac{i}{\hbar}[\bm{\Omega}_{\bf  k}\cdot{\bgreek\sigma},\rho_{\mu
  {\bf k}}(t)]-\frac{2\pi}{\hbar}\sum_{{\bf k}^\prime} M_{\bf
  k-k^\prime}I_{\bf kk^\prime}\\
&&\mbox{}\times\delta(\varepsilon_{\bf
  k}-\varepsilon_{\bf k^\prime})[\rho_{\mu {\bf k}}(t)-\rho_{\mu {\bf
    k^\prime}}(t)].
\label{a1}
\end{eqnarray}
Here $M_{\bf k-k^\prime}=|U^s_{\bf k-k^\prime}|^2+|U^{\rm Au}_{\bf
  k-k^\prime}|^2$ is the total electron-impurity scattering
matrix element contributed by impurities in the substrate and Au adatoms. $I_{\bf kk^\prime}=\frac{1}{2}[1+\cos(\theta_{\bf
    k}-\theta_{\bf k^\prime})]$ is the form factor.\cite{yzhou} By
expanding $\rho_{\mu\bf k}(t)$ as $\rho_{\mu\bf
  k}(t)=\sum_{l}\rho_{\mu k}^l(t)e^{il\theta_{\bf k}}$, one comes to
\begin{eqnarray}\nonumber
\partial_t\rho^l_{\mu k}(t)=&&-\frac{\alpha_{\rm R}}{2\hbar}[\sigma_+,\rho_{\mu
  k}^{l+1}(t)]+\frac{\alpha_{\rm R}}{2\hbar}[\sigma_-,\rho_{\mu k}^{l-1}(t)]\\ &&\mbox{}-\frac{\rho_{\mu
    k}^{l}(t)}{\tau^l_k},
\end{eqnarray}
where $\sigma_{\pm}=\sigma_x\pm i\sigma_y$, and  
\begin{equation}
\frac{1}{\tau^l_k}=\frac{k(1-\delta_{l0})}{4\pi\hbar^2v_{\rm
    F}}\int_0^{2\pi}d\theta
M_{\bf q}(1+\cos\theta)(1-\cos
l\theta)
\label{taukl}
\end{equation}
with $M_{\bf q}$ depending only on $|{\bf q}|=2k\sin\frac{\theta}{2}$. It is
noted that $\frac{1}{\tau^l_k}=\frac{1}{\tau^{-l}_k}$. 

Retaining the lowest three orders of $\rho_{\mu k}^l(t)$, i.e., $l=0$, $\pm 1$, and using the initial
conditions $\rho_{\mu k}^l(0)=\delta_{l0}\rho_{\mu k}^0(0)$, one obtains the second-order differential equation about $\rho_{\mu
  k}^0(t)$ as
\begin{eqnarray}\nonumber
 &&\partial^2_t\rho^0_{\mu k}(t)+\frac{1}{\tau_k^1}\partial_t\rho_{\mu
  k}^0(t)+\frac{\alpha_{\rm R}^2}{2\hbar^2}[\sigma_x, [\sigma_x,
     \rho_{\mu k}^0(t)]]\\&&\mbox{}+\frac{\alpha_{\rm R}^2}{2\hbar^2}[\sigma_y,
   [\sigma_y, \rho_{\mu k}^0(t)]]=0
\label{relax}
\end{eqnarray}
with  an affiliated initial condition $\partial_t\rho_{\mu
  k}^0(0)=0$. Defining the spin vector as ${\bf S}_{\mu
  k}^0(t)=\mbox{Tr}[\rho_{\mu k}^0(t){\bgreek \sigma}]$, one can obtain an equation
satisfied by ${\bf S}_{\mu  k}^0(t)$ directly from the above one,
which reads
\begin{eqnarray}
\Big[\partial_t^2+\frac{1}{\tau_k^1}\partial_t+\frac{2\alpha_{\rm
    R}^2}{\hbar^2}(1+\delta_{\alpha z})\Big]S_{\mu k\alpha}^0(t)=0 
\end{eqnarray}
with $\alpha=x,y,z$. With the initial condition $\partial_tS_{\mu
  k\alpha}^0(0)=0$, $S_{\mu  k\alpha}^0(t)$ is solved to be 
\begin{eqnarray}\nonumber
S_{\mu  k\alpha}^0(t)=&&\frac{S_{\mu
    k\alpha}^0(0)}{2}\Big[\Big(1+\frac{1}{\sqrt{1-c_\alpha^2}}\Big)e^{-\frac{t}{2\tau_k^1}(1-\sqrt{1-c_\alpha^2})}\\
  &&\mbox{}+\Big(1-\frac{1}{\sqrt{1-c_\alpha^2}}\Big)e^{-\frac{t}{2\tau_k^1}(1+\sqrt{1-c_\alpha^2})}\Big],
\end{eqnarray}
where $c_\alpha=2\sqrt{2(1+\delta_{\alpha z})}\alpha_{\rm
  R}\tau_k^1/\hbar$. When the scattering is strong enough and hence
$c_\alpha\ll 1$, 
\begin{eqnarray}\nonumber
  S_{\mu  k\alpha}^0(t)&\approx& S_{\mu
    k\alpha}^0(0)e^{-\frac{t}{4\tau_k^1/c_\alpha^2}}\\
  &\equiv&S_{\mu k\alpha}^0(0)e^{-\frac{t}{\tau_\alpha}}.
\label{a7}
\end{eqnarray}

As a result, for spins polarized along the $x$- and $y$-axis the spin relaxation times are
$\tau_x=\tau_y=\hbar^2/(2\alpha_{\rm R}^2\tau_k^1)$, while for spins
polarized along the $z$-axis
$\tau_z=\hbar^2/(4\alpha_{\rm R}^2\tau_k^1)$. From Eq.~(\ref{taukl}) one
notices that $\tau_k^1$ is in fact the momentum relaxation time
$\tau_p(k)$. For the highly degenerate electron system in
graphene, $\tau_p(k)\approx\tau_p(k_{\rm F})\approx \tau_p$. Therefore
we have $\tau_x=\tau_y=2\tau_z=\hbar^2/(2\alpha_{\rm
  R}^2\tau_p)$. 

\section{Spin diffusion in graphene}
\label{appb}
The spin diffusion in the absence of an electric field is also investigated for the case with only the
electron-impurity scattering included. Performing angle expansion on
the steady-state KSBEs in a way similar to that shown in
Appendix~\ref{appa}, one arrives at
\begin{eqnarray}\nonumber
&&\partial_x\sum_{l_0=\pm 1}\rho_{\mu k}^{l+l_0}(x)+\gamma[\sigma_+,\rho_{\mu
      k}^{l+1}(x)] -\gamma[\sigma_-,\rho_{\mu
  k}^{l-1}(x)]\\ &&\mbox{}+\frac{2}{v_{\rm F}}\frac{\rho_{\mu k}^{l}(x)}{\tau_k^l}=0
\end{eqnarray}
with $\gamma=\alpha_{\rm R}/(\hbar v_{\rm F})$. Retaining the lowest
three orders of $\rho_{\mu k}^l(x)$ one obtains three equations involving $\rho_{\mu 
  k}^{0,\pm 1}(x)$ as 
\begin{eqnarray}
&&\hspace{-1 cm}\partial_x\sum_{l_0=\pm 1}\rho_{\mu k}^{l_0}(x)+\gamma[\sigma_+,\rho_{\mu
      k}^{1}(x)]-\gamma[\sigma_-,\rho_{\mu
  k}^{-1}(x)]=0,\label{l0}\\
&&\hspace{-1 cm}\partial_x\rho_{\mu k}^{0}(x)-\gamma[\sigma_-,\rho_{\mu
      k}^{0}(x)]+\frac{2}{v_{\rm F}}\frac{\rho_{\mu
      k}^{1}(x)}{\tau_k^1}=0,\label{l1}\\
&&\hspace{-1 cm}\partial_x\rho_{\mu k}^{0}(x)+\gamma[\sigma_+,\rho_{\mu
      k}^{0}(x)]+\frac{2}{v_{\rm F}}\frac{\rho_{\mu
      k}^{-1}(x)}{\tau_k^1}=0.
\label{l-1}
\end{eqnarray}
From these equations one immediately arrives at 
Eq.~(\ref{dif}) with $\tau_k^1$ being irrelevant. By multiplying
${\bgreek \sigma}$ and performing trace on both sides of
Eq.~(\ref{dif}), one gets the equation satisfied by ${\bf 
S}_{\mu k}^0(x)$ which can be written as
\begin{eqnarray}
\hspace{-1.2 cm }
&&\left(\begin{array}{ccc}
  \partial_x^2-4\gamma^2 & 0 &  -4\gamma\partial_x \\
  0 & \partial_x^2-4\gamma^2 & 0 \\
  4\gamma\partial_x & 0 &  \partial_x^2-8\gamma^2
\end{array}\right)\left(\begin{array}{c} S_{\mu kx}^0(x) \\S_{\mu
    ky}^0(x)\\S_{\mu kz}^0(x)\end{array}\right)=0.
\label{seq1}
\end{eqnarray}
With specified boundary conditions, ${\bf S}_{\mu k}^0(x)$ is solved and the
total spin signal ${\bf S}(x)$ is obtained by Eq.~(\ref{sum}), as
presented in Sec.~\ref{sdanalytical}. Explicitly, taking the boundary
condition (I) given in Sec.~\ref{sdanalytical} as an example, one obtains ${\bf S}_{\mu k}^0(x)$ as 
\begin{eqnarray}
\hspace{-1.2 cm}&&{\bf S}_{\mu {\bf k}}^0(x)=S_{\mu
  k}^0(0)e^{-x/l_x}\left(\begin{array}{c}\sqrt{1+\Delta^2}\sin(\omega
  x+\phi) \\0\\ c_1\sin(\omega x)\end{array}\right),
\end{eqnarray}
with the parameters $l_x$, $\omega$, $\Delta$, $\phi$ and $c_1$ given
in Sec.~\ref{sdanalytical}. By further summing over ${\bf k}$ and
$\mu$ one arrives at Eq.~(\ref{sx}).

\section{Spin transport in graphene}
\label{appc}
The analytical study of spin transport is carried out analogly. The
driving term from the electric field in the steady state is
approximated as
\begin{eqnarray}\nonumber
\frac{eE}{\hbar}\frac{\partial \rho_{\mu {\bf k}}(x)}{\partial
  k_x}&=&\frac{eE}{\hbar}\frac{\partial \rho_{\mu {\bf
      k}}(x)}{\partial \varepsilon_{\bf k}}\frac{\partial
 \varepsilon_{\bf k}}{\partial k_x}\\\nonumber
\\  &\approx& eEv_{\rm F}\cos\theta_{\bf k}\frac{\partial \rho_{\mu k}^0(x)}{\partial \varepsilon_{\bf k}}.
\end{eqnarray}
Then the Fourier transformation of the steady-state KSBEs reads
\begin{eqnarray}\nonumber
&&\partial_x\sum_{l_0=\pm 1}\rho_{\mu k}^{l+l_0}(x)+\gamma[\sigma_+,\rho_{\mu
      k}^{l+1}(x)] -\gamma[\sigma_-,\rho_{\mu
  k}^{l-1}(x)]\\ &&\mbox{}-eE\frac{\partial \rho_{\mu
      k}^0(x)}{\partial \varepsilon_{\bf k}}(\delta_{l-1}+\delta_{l1})+\frac{2}{v_{\rm F}}\frac{\rho_{\mu
      k}^{l}(x)}{\tau_k^l}=0.
\end{eqnarray}
From this equation one comes to Eq.~(\ref{transport}) by retaining the
lowest three orders of $\rho_{\mu k}^l(x)$. The equation satisfied by ${\bf 
S}_{\mu k}^0(x)$ is then 

\begin{widetext}

\begin{eqnarray} 
&&\left(\begin{array}{ccc}
  \partial_x^2-eE\partial_x\partial_{\varepsilon_{\bf k}}-4\gamma^2 & 0 &  -4\gamma\partial_x+2eE\gamma\partial_{\varepsilon_{\bf k}} \\
  0 & \partial_x^2-eE\partial_x\partial_{\varepsilon_{\bf k}}-4\gamma^2 & 0 \\
  4\gamma\partial_x-2eE\gamma\partial_{\varepsilon_{\bf k}} & 0 &  \partial_x^2-eE\partial_x\partial_{\varepsilon_{\bf k}}-8\gamma^2
\end{array}\right)\left(\begin{array}{c} S_{\mu kx}^0(x) \\S_{\mu
      ky}^0(x)\\S_{\mu kz}^0(x)\end{array}\right)=0.
\label{seq2}
\end{eqnarray}
\end{widetext}
Having the experience of solving Eq.~({\ref{seq1}}), we assume that ${\bf S}_{\mu k}^0(x)$ has the solution as ${\bf S}^0_{\mu
  k}(x)=S_{\mu k}^0(0){\bf T}(x)$ and therefore ${\bf S}(x)=S(0){\bf
  T}(x)$. Performing summation over $\mu$ and
${\bf k}$ on both sides of the above equation and using the trick
\begin{eqnarray}\nonumber
\int_0^{+\infty}dkk[\partial_{\varepsilon_{\bf k}}S_{\mu k}^0(0)]{\bf
  T}(x)&=&-\frac{\int_0^{+\infty}d\varepsilon_{\bf
    k}S^0_{\mu k}(0)}{\hbar^2v_{\rm F}^2}{\bf T}(x)
\\ 
&=&-\frac{\frac{1}{\beta}\ln\frac{1+e^{\beta\mu_{\uparrow}}}{1+e^{\beta\mu_{\downarrow}}}}{S(0)\hbar^2v_{\rm F}^2}{\bf S}(x),
\end{eqnarray}
one obtains the equation satisfied by ${\bf S}(x)$ as 

\begin{eqnarray} \nonumber
\hspace{-0.5 cm}&&\left(\begin{array}{ccc}
  \partial_x^2+\gamma{\mathcal E}\partial_x-4\gamma^2 & 0 &  -4\gamma\partial_x-2\gamma^2{\mathcal E} \\
  0 & \partial_x^2+\gamma{\mathcal E}\partial_x-4\gamma^2 & 0 \\
  4\gamma\partial_x+2\gamma^2{\mathcal E} & 0 &  \partial_x^2+\gamma{\mathcal E}\partial_x-8\gamma^2
\end{array}\right)\\\hspace{-0.5 cm}\mbox{} &&\times\left(\begin{array}{c} S_x(x) \\S_y(x)\\S_z(x)\end{array}\right)=0,
\label{seq3}
\end{eqnarray}
in which ${\mathcal E}$ is given by Eq.~(\ref{chi}). With the three different typical
boundary conditions presented in Sec.~\ref{sdanalytical}, ${\bf S}(x)$
is solved to have the same form as Eqs.~(\ref{sx})-(\ref{sz}) except
that the parameters are now given in Sec.~\ref{stanalytical}. 

We now calculate ${\mathcal E}$ in both the degenerate and nondegenerate
limits. In the degenerate limit,
$\ln\frac{1+e^{\beta\mu_{\uparrow}}}{1+e^{\beta\mu_{\downarrow}}}\approx\ln e^{\beta(\varepsilon_{{\rm F}\uparrow}-\varepsilon_{{\rm
      F}\downarrow})}=\beta\hbar v_{\rm F}(k_{{\rm F}\uparrow}-k_{{\rm
      F}\downarrow})=\beta\hbar v_{\rm F}\sqrt{\pi
  N_e}(\sqrt{1+P_0}-\sqrt{1-P_0})\approx\beta\hbar v_{\rm F}\sqrt{\pi
  N_e}P_0$. Making use of the relation $S(0)=N_eP_0$, one has 

\begin{eqnarray}
{\mathcal E}
&\approx&\frac{eE}{\sqrt{\pi N_e}\alpha_{\rm  R}}=\frac{eE}{\alpha_{\rm   R}k_{\rm F}},
\end{eqnarray}
where $k_{\rm F}=\sqrt{\pi N_e}$, the magnitude of Fermi
momentum of unpolarized electrons with density being $N_e$. In the
nondegenerate limit, $\ln\frac{1+e^{\beta\mu_{\uparrow}}}{1+e^{\beta\mu_{\downarrow}}}\approx
e^{\beta\mu_{\uparrow}}-e^{\beta\mu_{\downarrow}}$ and 
\begin{eqnarray}\nonumber
S(0)&\approx&\frac{1}{\pi}\int_0^{+\infty}dkk[e^{-\beta(\varepsilon_{\bf
      k}-\mu_{\uparrow})}-e^{-\beta(\varepsilon_{\bf
      k}-\mu_{\downarrow})}]\\
&=&\frac{1}{\pi(\beta\hbar v_{\rm F})^2}(e^{\beta\mu_{\uparrow}}-e^{\beta\mu_{\downarrow}}),
\end{eqnarray}
therefore 
\begin{equation}
{\mathcal E}\approx \frac{eE\beta\hbar v_{\rm   F}}{\alpha_{\rm R}}.
\end{equation}

\section{Derivation of two-component drift-diffusion equation 
from KSBEs}
\label{dde}

The two-component drift-diffusion equation can be derived from the
KSBEs in the collinear spin space\cite{cheng083704} with
the $z$-axis along the initial spin-polarization direction ${\bf\hat n}$, by
neglecting the spin coherence (i.e., the off-diagonal components
 of the density matrices). The density matrices then have
the diagonal form as $\frac{1}{2}[f_{\mu{\bf k}\uparrow}(x,t)+f_{\mu{\bf
    k}\downarrow}(x,t)+(f_{\mu{\bf k}\uparrow}(x,t)-f_{\mu{\bf
    k}\downarrow}(x,t))\sigma_z]$. In the following we present a brief
derivation of the two-component drift-diffusion equation from the
KSBEs with only the strong electron-impurity scattering 
considered. Other kinds of scattering can also be incorporated similarly
under elastic scattering approximation. The spin relaxation time is 
$\tau_{\bf\hat n}$ 
and the momentum relaxation time is $\tau_p$ (both are given in
Appendix~\ref{appa}) and we neglect their momentum dependence
hereafter. The two-component drift-diffusion equation is obtained from 
the equation of continuity and the equation of 
current, both to be derived from the KSBEs. 

The equation of continuity is derived as follows. By multiplying the
 KSBEs [Eq.~(\ref{ksbee})] with
$\frac{1}{2}(1\pm\sigma_z)$ and then performing the trace,
one obtains the simplified KSBEs for each spin band
($\sigma=\uparrow$, $\downarrow$) as 
\begin{eqnarray}\nonumber
&&\frac{\partial f_{\mu{\bf k}\sigma}(x,t)}{\partial
  t}-\frac{eE}{\hbar}\frac{\partial f_{\mu{\bf  k}\sigma}(x,t)}{\partial
  k_x}+v_{\rm F}\cos\theta_{\bf k}\frac{\partial f_{\mu{\bf  k}\sigma}(x,t)}{\partial
  x}\\ &&=-\frac{f_{\mu{\bf k}\sigma}(x,t)-f_{\mu{\bf
      k}-\sigma}(x,t)}{2\tau_{\bf \hat n}}.
\label{d1}
\end{eqnarray} 
The right-hand side of the above equation comes from the term
$\mbox{Tr}\{\frac{1}{2}(1\pm\sigma_z)[\partial_t\rho_{\mu\bf k}(x,t)|_{\rm coh}+\partial_t\rho_{\mu\bf
  k}(x,t)|_{\rm scat}]\}$, which can be calculated with the aid of the
 KSBEs in the time domain
[Eq.~(\ref{a1})] and the corresponding 
solution [Eq.~(\ref{a7})]. Performing summation over $\mu$ and ${\bf k}$ 
on both sides of Eq.~(\ref{d1}) in the steady state, one comes to
\begin{eqnarray}\nonumber
&& -\frac{eE}{\hbar}\sum_{\mu{\bf k}}\frac{\partial f_{\mu{\bf  k}\sigma}(x)}{\partial
  k_x}+\frac{\partial}{\partial x}\sum_{\mu\bf k}v_{\rm
  F}\cos\theta_{\bf k}f_{\mu{\bf  k}\sigma}(x)\\ 
&& =-\frac{N_\sigma(x)-N_{-\sigma}(x)}{2\tau_{\bf\hat n}},
\end{eqnarray}
where $N_\sigma(x)$ is the electron density with spin $\sigma$ at
position $x$. Up to the first order of the electric field $E$, 
the first summation over ${\bf k}$ in the above
equation leads to zero when $f_{\mu{\bf k}\sigma}$ is approximated 
by $f_{\mu{\bf k}\sigma}^0$, the distribution in equilibrium. 
 Defining the charge current along the $x$-axis with spin $\sigma$ as 
\begin{equation}
J_\sigma(x)=-\sum_{\mu{\bf k}}ev_{x}f_{\mu{\bf
    k}\sigma}(x)
\end{equation}
with $v_{x}=v_{\rm F}\cos\theta_{\bf k}$, one has the equation
of continuity
\begin{equation}
-\frac{1}{e}\frac{\partial J_\sigma(x)}{\partial
  x}=-\frac{N_\sigma(x)-N_{-\sigma}(x)}{2\tau_{\bf\hat n}}.
\label{d2}
\end{equation}

We then calculate the current $J_\sigma$ from the diagonal part of the
KSBEs 
\begin{eqnarray}\nonumber
&&\frac{\partial f_{\mu{\bf k}\sigma}(x,t)}{\partial
  t}-\frac{eE}{\hbar}\frac{\partial f_{\mu{\bf  k}\sigma}(x,t)}{\partial
  k_x}+v_{\rm F}\cos\theta_{\bf k}\frac{\partial f_{\mu{\bf  k}\sigma}(x,t)}{\partial
  x}\\ &&=-\frac{f_{\mu{\bf k}\sigma}(x,t)-f_{\mu{\bf
      k}\sigma}^0(x,t)}{\tau_p}, 
\end{eqnarray}
where the right-hand side of the equation comes from the
electron-impurity scattering (Appendix~\ref{appa}). In the steady state,
multiplying $-ev_{x}$ on both sides of the equation and then 
summing over $\mu$ and ${\bf k}$, one comes to
\begin{eqnarray}\nonumber
&& \frac{e^2Ev_{\rm F}}{\hbar}\sum_{\mu\bf k}\cos\theta_{\bf
  k}\frac{\partial f_{\mu{\bf  k}\sigma}(x)}{\partial
  k_x}-ev_{\rm F}^2\sum_{\mu\bf k}\cos^2\theta_{\bf k}\frac{\partial f_{\mu{\bf  k}\sigma}(x)}{\partial
  x}\\ && =-\frac{J_\sigma(x)}{\tau_p}.
\end{eqnarray}
Again, up to the first order of the electric field, one has
\begin{equation}
J_\sigma(x)=e\mu_\sigma EN_\sigma(x)+eD\partial_x N_\sigma(x),
\label{d3}
\end{equation}
where the mobility $\mu_\sigma=\frac{ev_{\rm
    F}\tau_p}{\hbar\sqrt{2\pi N_\sigma}}$ and the charge diffusion
coefficient $D=\frac{1}{2}v_{\rm F}^2\tau_p$. For the case with small spin
polarization, $\mu_\uparrow\approx\mu_\downarrow=\mu_e\equiv\frac{ev_{\rm
    F}\tau_p}{\hbar\sqrt{\pi N_e}}$ in which
$N_e=N_\uparrow+N_\downarrow$ is the total electron density.

Finally, the two-component drift-diffusion equation is obtained by combining
Eqs.~(\ref{d2}) and (\ref{d3})  
\begin{equation}
 -\mu_eE\frac{\partial N_\sigma(x)}{\partial x}-D\frac{\partial^2
  N_\sigma(x)}{\partial x^2}=-\frac{N_\sigma(x)-N_{-\sigma}(x)}{2\tau_{\bf\hat n}}.
\end{equation}
This equation is consistent with that in the
literature.\cite{yu,fabiandd,fabiandd1,fabiandd2} 
The equation of $\Delta N=N_\uparrow-N_\downarrow$ then reads
\begin{equation}
 -\mu_eE\frac{\partial \Delta N(x)}{\partial x}-D\frac{\partial^2
  \Delta N(x)}{\partial x^2}=-\frac{\Delta N(x)}{\tau_{\bf\hat n}}.
 \end{equation}
With boundary condition $\Delta N(+\infty)=0$,
$\Delta N(x)$ is solved as $\Delta N(x)=\Delta N(0)e^{-x/l_{\bf\hat
  n}}$ where the spin transport length\cite{yu}
\begin{equation}
l_{\bf\hat
  n}=\left [\frac{\mu_eE}{2D}+\sqrt{\big(\frac{\mu_eE}{2D}\big)^2+\frac{1}{D\tau_{\bf\hat
    n}}}\right ]^{-1}.
\label{d5}
\end{equation}
When the electric field is zero,
\begin{equation}
l_{\bf\hat
  n}=\sqrt{D\tau_{\bf \hat n}}.
\label{d6}
\end{equation}
As indicated by Eqs.~(\ref{d5}) and
(\ref{d6}), in the frame of the two-component drift-diffusion model,
the anisotropy of the spin transport or diffusion is solely determined 
by that of the spin relaxation and no spatial spin
precession can be obtained.
\end{appendix}

\end{document}